\documentclass{aa}
\begin{document}
\title{Two-fluid matter-quintessence FLRW models:
energy transfer and the equation of state of the universe}
\author{A.~Gromov\inst{1,4},
Yu.~Baryshev\inst{2,4},
P.~Teerikorpi\inst{3}
}
\offprints{}
\institute{
Centre for Advanced Studies of the Saint-Petersburg State Technical
University,
195251, Politechnicheskaya 29, St. Petersburg, Russia
\and
 Astronomical Institute of the Saint-Petersburg University,
198904 St. Petersburg, Russia
\and
Tuorla Observatory, FIN-21500 Piikki\"{o}, Finland,
\and
Isaac Newton Institute of Chile, St.Petersburg Branch
}
\date{Received \hspace*{3cm}; accepted}

\abstract{Recent observations support the view
that the universe is described by a FLRW model with
$\Omega_m^0 \approx 0.3$, $\Omega_{\Lambda}^0 \approx 0.7$, and $w \leq
-1/3$ at the present epoch.
There are several theoretical suggestions for the
cosmological $\Lambda$ component and for the
 particular form
of the energy transfer between this dark energy and matter.
 This gives a strong motive for a systematic study of general properties of
two-fluid FLRW models. We consider a combination of one perfect
fluid, which is quintessence with negative pressure
($p_Q = w\epsilon_Q$ ), and another perfect fluid,
 which is a mixture of
radiation and/or matter components with positive pressure
($p = \beta \epsilon_m$), which define the associated one-fluid model
($p = \gamma \epsilon$).
 We introduce a useful classification which contains 4 classes of
models defined by the presence or absence of energy transfer and
by the stationarity ($w = const.$ and $\beta = const.$) or/and
non stationarity ($w$ or $\beta$ time dependent)
of the equations of state.
 It is shown
that, for given $w$ and $\beta$, the
energy transfer defines $\gamma$ and, therefore, the total
gravitating mass and dynamics of the model. We study
important  examples of two-fluid FLRW models within the new classification.
The behaviour of the energy content, gravitating mass, pressure, and
the energy transfer are given as functions of
the scale factor.
We point out
three characteristic scales, $a_E$, $a_{\cal P}$ and $a_{\cal M}$,
which separate periods of time in which quintessence energy, pressure
and gravitating mass dominate. Each sequence of the scales defines
one of 6 evolution types.
 }
\maketitle

\section{Introduction} \label{introd}

A number of
recent observations reveal
the cosmological $\Lambda$ component. Type Ia Supernovae
(Riess et al. \cite{Riess98}, Perlmutter et al.
\cite{Perl}, Riess et al. \cite{Riess01}) and the Boomerang,
Maxima and Dasi measurements of the total density parameter
$\Omega$ via the first acoustic peak location in the angular
power spectrum of the CBR
(de Bernardis et al. \cite{deBern00}, Balbi et al. \cite{Balbi00},
Jaffe et al. \cite{Jaffe00}) show that
$\Omega = \Omega_m^0 +\Omega_{\Lambda}^0 = 1 \pm 0.05$ with
$\Omega_{\Lambda}^0 \approx 0.7$.
Existing cosmological data allow a wide
range for the equation of state coefficient $w$
 from $-1$ to $-1/3$ (Perlmutter et al.
\cite{PerlTur99}; Podariu \& Ratra \cite{Pod00}; Wang et al.
\cite{Wang00}).

Also, the smooth Hubble flow around our Local Group,
inside a highly lumpy matter distribution, suggests still other
evidence for a dominating $\Lambda$ component and its variation
with time
(Chernin et al. \cite{CherTerBar00}; Baryshev et al.
\cite{BarCherTer01}; Chernin \cite{CherninUFN};
Klypin et al. \cite{Klypin01}; Axenides \& Perivolaropoulos
\cite{Axenides02}).

There are several theoretical models for $\Lambda$-like
cosmological components of the universe with positive energy
density and negative pressure,
including vacuum with a constant $\Lambda$, decaying $\Lambda$, and
variable equation of state $w(t)$
(Peebles \& Ratra \cite{PeebRat88};
Lima \& Maia \cite{Lima94};
 Wetterich \cite{Wett95};
Ferreira \&
Joyce  \cite{Ferr97}; Caldwell et al.
\cite{Cald98}; Steinhardt et al. \cite{Stein98};
Zlatev et al. \cite{Zlatev99};
Bahcall et al. \cite{bahcall99}; Mbonye \cite{Mbonye02}).
We use the term ``quintessence'' for
any kind of substance having the equation of state $p_Q =
w\,\varepsilon_Q$ with $-1 \le w < 0$, which may be time variable.

Thus observations and theory give strong motivation to study
general properties of
two-fluid Friedmann--Lema\^{i}tre--Robertson--Walker (FRLW)
models in which the $\Lambda$ component dominates at late epochs
and there is energy transfer between $\Lambda$ and matter
components.

The energy transfer between dust-like matter and radiation was first studied
by Davidson (\cite{David62}). Some examples of the physics
producing the energy transfer were given in Sistero (\cite{Sis71})
and McIntosh (\cite{McIn1967}, \cite{McIn1968}). The first exact
solution of the equation of motion for a dust+radiation model with
no energy transfer was obtained by Chernin (\cite{Chernin65}), who
applied it to the case where the radiation  component is the
cosmic background radiation or neutrinos.

In the present paper we give a systematic presentation of the
properties of two-fluid cosmological models, with and without
energy transfer, in the frame of a
new classification which naturally arises in the two-fluid problem.
In section 2 we briefly summarize the standard two-fluid FLRW model.
In section 3 we consider general properties of two-fluid models
with matter and quintessence. Section
4 includes three examples of models from different
classes of our classification. Section 5 contains the conclusions.

\section{The two-fluid model: a summary}

The derivation of the FLRW equations contains the following basic
elements: 1) The Einstein equations (notations from
Landau \& Lifshitz \cite{LL71})
\begin{eqnarray}
\Re_k^i - \frac{1}{2}g_k^i\,\Re = \frac{8\pi G}{c^4}T_{k}^{i}
\label{einstein}
\end{eqnarray}
2) The Bianchi identity in the form:
\begin{eqnarray}
{T_{k}^{i}}_{;i}
= 0
\label{divergence}
\end{eqnarray}
3) The RW line element in spherical comoving space coordinates
($\chi, \theta, \phi$) and synchronous time $t$:
\begin{eqnarray}
ds^2 = c^2dt^2 - S^2(t)d\chi^2 - S^2(t) I_k^2 (\chi)
(d\theta^2 + sin^2 \theta d\phi^2)
\label{RW}
\end{eqnarray}
4) The total energy momentum tensor
in comoving coordinates (ordinary matter, vacuum, quintessence):
\begin{eqnarray}
T_{k}^{i} = diag(\varepsilon,-p,-p,-p).
\label{EMT}
\end{eqnarray}
Here
$I_k(\chi) = (sin(\chi),~\chi,~sinh(\chi))$ for the curvature
constant $k = (+1,0,-1)$, respectively. $S(t)$ is the scale
factor, $\varepsilon = \rho c^2$ is the energy density, and $p$
is the pressure.

The proper metric distance $r$ of a particle with the dimensionless
comoving coordinate $\chi$ from the observer is
\begin{eqnarray}
r(t, \chi) = S(t) \cdot \chi.
\label{r}
\end{eqnarray}
Here physical dimensions $[r]=[S]=$ cm.
 The volume element in metric (\ref{RW}) $dV =
S^3\,I_k^2(\chi)\,sin^2(\theta) \,d \chi\, d \theta\, d \phi$.

\subsection{The FLRW one-fluid model} \label{FLRW 1fm}

For the one-fluid model the (0,0) and
(1,1) components of Eq.(\ref{einstein}) give the FLRW equations
\begin{eqnarray}
\frac{\dot S^2}{S^2} + \frac{kc^2}{S^2} = \frac{8\pi
G}{3c^2}\varepsilon,
\label{0.0}
\end{eqnarray}
\begin{eqnarray}
2\,\frac{\ddot S}{S} + \frac{\dot S^2}{S^2}
+ \frac{kc^2}{S^2}
= -\frac{8\pi G}{c^2}\,p,
\label{1.1}
\end{eqnarray}
where $\dot{} = d/dt$. Eq.(\ref{divergence}) implies:
\begin{eqnarray}
3 \frac{\dot S}{S} = -\frac{\dot \varepsilon}
{\varepsilon + p}.
\label{continuity b}
\end{eqnarray}
The equation of state for the one-fluid model is
\begin{eqnarray}
p = \gamma\,\varepsilon.
\label{uuu}
\end{eqnarray}
The characteristic energy content $e$ of the sphere with the
radius equal to the proper metric distance $r= S(t)\chi$ is:
%
\begin{eqnarray}
e(r) = \int
\varepsilon dV =
4\pi\,\varepsilon\,S^3\,\sigma_k(\chi),
\label{e content 1fm}
\end{eqnarray}
where for different values of $k = +1, 0, -1$, $\,$
$\sigma_k(\chi) =
 \frac{\chi^2}{2} - \frac{\sin 2\chi}{4}$,
$\frac{\chi^3}{3}$,
$\frac{\sinh 2\chi}{4}-\frac{\chi^2}{2}$, respectively.

The gravitating mass $M_g$
is defined by the equation of motion, which follows from
Eqs.(\ref{0.0}) - (\ref{1.1}):
\begin{eqnarray}
\ddot r = - G\,\frac{M_g(r)}{r^2},
\label{eq mot}
\end{eqnarray}
where
\begin{eqnarray}
M_g(r) = \frac{\varepsilon + 3p}{c^2}\int\limits_{0}^{r}dV =
\frac{e}{c^2}\,(3\,\gamma + 1).
\label{mgrav}
\end{eqnarray}

This definition corresponds to the general expression for the gravitating
mass of a
self-gravitating body as given by Landau \& Lifshitz \cite{LL71},
\S 100).

\subsection{Time dependence of energy}

There are two factors which cause a time dependence of energy of
each particular
fluid and, hence, of an associated (as termed by Coley \&
Wainwright \cite{CW92}) one-fluid model. The first cause
is the expansion of the universe. The second is when one
fluid is converted  into another.

It is known (Lema{\^{i}}tre \cite{L1933}; Davison
\cite{David62};
 Harrison \cite{Harr95}; Peebles \cite{Peebl.book93}, p.139) that
Eqs.(\ref{0.0}) and (\ref{1.1}) imply  the non-conservation of the
energy $\varepsilon$ of one-fluid FLRW models filled
by a non dust-like matter:
\begin{eqnarray}
\frac{d\,\varepsilon}{dt}
+ p\frac{d}{dt}(R^3) = 0.
\label{Dav}
\end{eqnarray}
For every particular fluid the time dependence of the energy is defined
by
its pressure. Eq.(\ref{Dav}) is a consequence of the Bianchi identity
(\ref{continuity b}).
Rewritten for each particular fluid,
it has the form of the classical energy conservation law
without the convertion of one-fluid to another:
\begin{eqnarray}
\frac{d\,\varepsilon_Q}{dt} + p_Q\frac{d}{dt}(R^3) = 0, \quad
\frac{d\,\varepsilon_m}{dt} + p_m\frac{d}{dt}(R^3) = 0.
\label{Dav 2fm}
\end{eqnarray}
When the one-fluid converts into another (for instance, the
matter into radiation in stars), Eqs.(\ref{Dav 2fm}) become
\begin{eqnarray}
\frac{1}{R^3}\left[
\frac{d\,\varepsilon_Q}{dt} + p_Q\frac{d}{dt}(R^3)\right] =
u_Q,\nonumber
\end{eqnarray}
\begin{eqnarray}
\frac{1}{R^3}\left[
\frac{d\,\varepsilon_m}{dt} + p_m\frac{d}{dt}(R^3)\right] =
u_m,\nonumber
\end{eqnarray}
\begin{eqnarray}
u_Q + u_m = 0,
\label{Dav 2fm U}
\end{eqnarray}
$u_m$ and $u_Q$ are the rates of energy transfer per unit
volume from the quintessence to matter  and vice versa
($[u_m]=$ erg/sec$\cdot$cm$^3$).
The terms $p_Q\frac{d}{dt}(R^3)$ and $p_m\frac{d}{dt}(R^3)$ describe
the time
dependence of the energy due to space expansion.

\subsection{The two-fluid model
 for matter and quintessence} \label{FLRW 2fm}

In his pioneering work, Lema\^{i}tre
(\cite{L1933}) studied a model containing dust and the
cosmological constant. The first application of the energy
transfer for two-fluid cosmology was by Davidson
(\cite{David62}). We also make use of the results
by Chernin
(\cite{Chernin65}), Thomas \& Schulz (\cite{Thom00}), Coley \&
Tupper (\cite{ColeyTupp86}), Amendola \& Tocchini--Valentini
(\cite{amendola00}).

We study two groups of substances:
1) a number of perfect fluids with positive
pressure; 2) a  number of fluids with positive energy and
negative pressure.
Each group plays the role of a one-fluid in two-fluid models.
The equations of state for the first group are:
\begin{eqnarray}
p_m = \beta\,\varepsilon_m,
\label{pm}
\end{eqnarray}
where $p_m = p_{m1} + p_{m2} + ...$,
$\varepsilon_m = \varepsilon_{m1} + \varepsilon_{m2} + ...$,
$0 \leq \beta_{i} \leq 1$ and $\beta = \frac{\varepsilon_m}{p_m} > 0$.
The second group has:
\begin{eqnarray}
p_Q = w\,\varepsilon_Q,
\label{pq}
\end{eqnarray}
where $p_Q = p_{Q1} + p_{Q2} + ...$,
$\varepsilon_Q = \varepsilon_{Q1} + \varepsilon_{Q2} + ...$,
$-1 \leq w_{i} < 0 $ and $w = \frac{\varepsilon_Q}{p_Q} < 0$.

We define an important function
\begin{eqnarray}
\alpha(a) = \frac{\varepsilon_Q(a)}{\varepsilon_m(a)},
\label{alpha}
\end{eqnarray}
which characterizes the relative contributions to the energy
density from quintessence and matter (see e.g.
Wetterich \cite{Wett95}).
 The relative energy contribution
was studied
in another way
 by Coley \& Wainwright (\cite{CW92}).

The two-fluid model is defined on two levels. Firstly, on a level
of partial one-fluid models, with two kinds of interactions, 1)
through the common gravitational field and 2) by the energy
transfer from one-fluid to the other. Secondly, on the level of
the associated one-fluid model.

For a complete description of a two-fluid model one has to study
both levels simultaneously. We assume state equations of unique
form for partial models and for the associated one-fluid model.
The essential difference is that the equation of state of the
associated model is not produced by physical particles
and their interactions.

\subsubsection{The associated one-fluid model}

We specify the partial energies, pressures and the corresponding
equations of state:
\begin{eqnarray}
\varepsilon_{Q} = \rho_{Q}c^2,\qquad
\varepsilon_m = \rho_m c^2,\qquad
\varepsilon = \varepsilon_{Q} + \varepsilon_{m},
\nonumber
\end{eqnarray}
\begin{eqnarray}
p_{Q} = w \varepsilon_{Q},\qquad
p_m = \beta \varepsilon_m, \qquad
p = p_{Q} + p_{m}, \nonumber
\end{eqnarray}
\begin{eqnarray}
p = \gamma \varepsilon.
\label{2 fluids}
\end{eqnarray}
Thus the equation of
state for the associated one-fluid is:
\begin{eqnarray}
\gamma = \frac{p}{\varepsilon} =
\frac{w\alpha + \beta}{\alpha + 1}.
\label{1fm eqs}
\end{eqnarray}
For the energy ratio $\alpha \ge 0$ one finds that $w \le \gamma
\le \beta$. The associated one-fluid description of the two-fluid
model includes Eqs.(\ref{0.0}) - (\ref{uuu}) and Eq.(\ref{1fm
eqs}). The three functions of
the scale factor $a$
 $w(a),\, \beta(a)$ and $\,\alpha(a)$
are assumed given.

\subsubsection{The two-fluid model}

The energy-momentum tensor of the associated one-fluid model is
the sum of the partial tensors of matter and quintessence.
According to Eq.(\ref{divergence}) we have
\begin{eqnarray}
{T_{k}^{i}}_{;i}
= T_{(m)k;i}^{~~~~i} +  T_{(Q)k;i}^{~~~~i} = 0;
\label{divergence 12 1}
\end{eqnarray}
Eq.(\ref{continuity b}) implies:
\begin{eqnarray}
3 \frac{\dot S}{S} = -\frac{\dot \varepsilon_{tot}}
{\varepsilon_{tot} + p_{tot}}, \nonumber
\end{eqnarray}
\begin{eqnarray}
\varepsilon_{tot} =\varepsilon_{Q} + \varepsilon_{m}, \qquad
p_{tot} = p_{Q} + p_{m}.
\label{continuity}
\end{eqnarray}

There are two subcases within the two-fluid model description. In
the first the energy-momentum tensors of the partial fluids are
separately conserved:
\begin{eqnarray}
T_{(m)k;i}^{~~~~i} = 0\qquad\mbox{and}\qquad
T_{(Q)k;i}^{~~~~i} = 0,
\label{divergence 12}
\end{eqnarray}
so that $u_{Q} = 0$ and $u_m = 0$.
In the second case we allow the presence of
the energy transfer,
when $u_Q \ne 0$ and $u_m \ne 0$ but $u_{Q} + u_m = 0$ which is
equivalent to Eqs.(\ref{continuity}).

Each fluid has an equation of state in the form of
Eq.(\ref{uuu}) with different coefficients: $\beta(a)$ for the
matter component and $w(a)$ for quintessence (see Eqs.(\ref{2
fluids}) ).

For two-fluid FLRW models the characteristic energy content within the
sphere of a fixed comoving radius $\chi$ is
\begin{eqnarray}
e = 4\,\pi\,\varepsilon S^3 \sigma_k(\chi) =
4\,\pi\,(\alpha + 1)\,\varepsilon_m\,S^3\,\sigma_k(\chi)
\label{e content}
\end{eqnarray}
The gravitating mass $M_g$ is the sum of the partial gravitating
masses of two-fluids:
\begin{eqnarray}
M_g(r) = M_g^m + M_g^{Q} =
\nonumber
\end{eqnarray}
\begin{eqnarray}
\frac{1}{c^2}\,\left[(1 + 3 w)\varepsilon_{Q}  + (1 +
3\beta)\varepsilon_m\,\right]\,\int\limits_{0}^{r}dV =
\frac{e}{c^2}\,(3\,\gamma + 1).
\label{MmL}
\end{eqnarray}

\subsection{Dimensionless equations}
\label{dimless}\label{sec3-1}

Now we restate the models in terms of dimensionless quantities.
We introduce the following characteristic values:
\begin{eqnarray}
t_0 = \frac{1}{H_0}, \quad  l_0 = c\,t_0 = R_{H_{0}},\quad
\rho_0 = \frac{3\,H^2_0}{8\,\pi\,G},
\end{eqnarray}
\begin{eqnarray}
M_0 = \frac{4\,\pi}{3}\,\rho_0\,R_{H_0}^3 =
\frac{c^2}{2\,H_0\,G},\qquad
\varepsilon_0 = \rho_0\,c^2,
\nonumber
\end{eqnarray}
and dimensionless variables:
\begin{eqnarray}
a(\tau) = \frac{S(t)}{l_0}, \quad
\tau = \frac{t}{t_0}, \quad \delta = \frac{\rho}{\rho_0},
\nonumber
\end{eqnarray}
\begin{eqnarray}
\mu(a) = \frac{M_g(t)}{M_0} = 3\sigma_k(\chi)
(3\gamma + 1)a^3{\cal E},
\nonumber
\end{eqnarray}
\begin{eqnarray}
{\cal E} &=& \frac{\varepsilon}{\varepsilon_0},\quad\quad
{\cal E}_{Q} = \frac{\varepsilon_Q}{\varepsilon_0},\quad\quad
~{\cal E}_{m} = \frac{\varepsilon_m}{\varepsilon_0},\quad\quad
{\cal E} = {\cal E}_{Q} + {\cal E}_{m}
\nonumber\\
P &=& \frac{p}{\varepsilon_0},\quad\quad
P_Q = \frac{p_Q}{\varepsilon_0},\quad\quad
P_m = \frac{p_m}{\varepsilon_0},\quad\quad
P = P_q + P_m
\nonumber\\
E &=& \frac{e}{e_0},\quad\quad
E_Q = \frac{e_Q}{e_0},\quad\quad
E_m = \frac{e_m}{e_0},\quad\quad
E = E_Q + E_m
\nonumber\\
U &=& \frac{u}{\varepsilon_0}\,t_0,\quad
U_Q = \frac{u_Q}{\varepsilon_0}\,t_0,\quad
U_m = \frac{u_m}{\varepsilon_0}\,t_0,\quad
U = U_Q + U_m
\label{d l v 2}
\nonumber
\end{eqnarray}
Eqs. (\ref{0.0}) - (\ref{uuu}) and (\ref{Dav 2fm U}) become
\begin{eqnarray}
\frac{\dot a^2}{a^2} + \frac{k}{a^2} = ~{\cal E},
\label{0.0 diml}
\end{eqnarray}
\begin{eqnarray}
2\,\frac{\ddot a}{a} + \frac{\dot a^2}{a^2} + \frac{k}{a^2} = -
3\,P,
\label{1.1 diml}
\end{eqnarray}
\begin{eqnarray}
3\,\frac{\dot a}{a} = -\frac{\dot {~{\cal E}}}{ ~{\cal E} + P},
\label{continuity dimless}
\end{eqnarray}
\begin{eqnarray}
P = \gamma\,~{\cal E},
\label{uuu dimless}
\end{eqnarray}
\begin{eqnarray}
U_{Q} = \frac{1}{a^3}\,\left[
\frac{d}{d\tau}\,\left(~{\cal E}_{Q}\,a^3 \right)
+ P_{Q}\,\frac{d}{d\tau}\,a^3 \right],
\label{E_Q dimless}
\end{eqnarray}
\begin{eqnarray}
U_m = \frac{1}{a^3}\,
\left[
\frac{d}{d\tau}\,\left(~{\cal E}_m\,a^3 \right) +
P_m\,\frac{d}{d\tau}\,a^3 \right].
\label{E_m dimless}
\end{eqnarray}
Eq.(\ref{eq mot}) for the scale factor now has the form:
\begin{eqnarray}
a^2\,\ddot a = - \Omega_0\,\mu(a),
\label{motion dl}
\end{eqnarray}
where $\Omega_0$ is the cosmological density parameter
 at the moment of the
initial conditions for Eq.(\ref{motion dl}).
\begin{table}
\caption{Classification of two-fluid FLRW models.\newline
ET means ``energy transfer''; NET is ``no energy transfer'';
SES is ``stationary equation of state'';
NSES is ``non stationary equation of state''.
}
\begin{center}
\begin{tabular}{|c|c|c|} \hline
                             &$U_Q = U_m = 0$    & $U_Q = -U_m
                             \ne 0$ \\ \hline
both $w$, $\beta$ &                             &          \\
 are constant                &NET-SES
 &ET-SES    \\  \hline
at least one of $w$, $\beta$  &                             &
\\
 is not constant                &NET-NSES
 &ET-NSES    \\  \hline
\end{tabular}
\end{center}
\label{epm 1}
\end{table}

\section{General properties of two-fluid FLRW models with matter and
quintessence} \label{classification equation}

\subsection{The classification} \label{classification}

A two-fluid model with the vacuum or quintessence together with some
mixture of matter is very different from the two-fluid
model with ``ordinary'' matter:  negative
pressure and gravitating mass components
give rise to a new behaviour of the total pressure and total
gravitating mass.
In order to facilitate their study,
we divide all two-fluid models into four classes according to two
independent properties: 1) both fluids have a  stationary (SES) or
at least one has a non stationary (NSES) equation of state and 2)
the presence (ET, energy transfer) or absence (NET, no energy
transfer) of an energy transfer between the components (see also
Table 1).

Furthermore, we separate three different kinds of two-fluid models
depending on the behaviour of the function $\alpha(a) = {\cal
E}_Q/{\cal E}_m$: the coherent model, when
\begin{eqnarray}
\alpha(a) = const,
\label{coh 1}
\end{eqnarray}
the asymptotically coherent model, when
\begin{eqnarray}
\lim_{a \to \infty}{\alpha^{\prime}(a)} = 0
\label{coh 2}
\end{eqnarray}
and the non coherent model, when neither Eq.(\ref{coh 1}) nor
Eq.(\ref{coh 2}) are valid.

\subsection{The general solution of two-fluid FLRW models:
ET-NSES models}

The ET-NSES class is the
most
general class of two-fluid models in our
classification. We assume that all fluids (the associated and the two
particular ones) have nonstationary equations of state.

In this subsection we completely describe the
two-fluid problem in terms of the associated
one-fluid model and two particular fluid models simultaneously. This
leads us to rewrite the FLRW equations in terms of the
coefficients of the three equations of state: $\gamma$, $\beta$
and $w$.

\subsubsection{The input equations and their solution}

We describe the associated one-fluid model by Eqs.(\ref{continuity
dimless},\ref{uuu dimless}):
\begin{eqnarray}
-3\,\frac{\dot a}{a} = \frac{\dot{~{\cal E}}}{~{\cal E} + P},
\qquad
P = \gamma\,~{\cal E},
\qquad
\gamma = \gamma(a).
\label{qwe 1}
\end{eqnarray}
The solution of Eqs.(\ref{qwe 1}), with the initial conditions stated
for the
present epoch, is:
\begin{eqnarray}
~{\cal E} = ~{\cal E}(1)\,\exp
\left(-3\,\int\limits_1^a \frac{\gamma + 1}{x}\,dx\right).
\label{def en den}
\end{eqnarray}
In this model the characteristic energy content $E$, pressure
$P$ and gravitating mass $\mu$ are:
\begin{eqnarray}
E = ~{\cal E}\,a^3, \quad P = ~{\cal E}\,\gamma,\quad
\mu = 3\,\sigma_k(\chi)\,{\cal E}\,a^3\,(3\,\gamma + 1).
\label{def a 11}
\end{eqnarray}

The scale factor satisfies Eq.(\ref{motion dl}). $\gamma(a)$ is
given by Eq.(\ref{1fm eqs}) in case of known $w(a)$ and
$\beta(a)$.

The two particular fluids are described by Eqs.(\ref{E_Q dimless})
- (\ref{E_m dimless}) and the equations of state:
\begin{eqnarray}
P_Q = w(a)\,~{\cal E}_Q, \quad
P_m = \beta(a)\,~{\cal E}_m
\label{E_Q 1}
\end{eqnarray}

Calculating $a^3\,\left( U_m - \frac{U_Q}{\alpha}\right)$ and
keeping in mind that $U_Q + U_m  = 0$, we find the equation
\begin{eqnarray}
\frac{U_m}{E_m}\,\frac{\alpha + 1}{\alpha}\,a^3 =
3\,\frac{\dot a}{a}\,(\beta -w) - \frac{\dot \alpha}{\alpha}.
\label{classific equation}
\end{eqnarray}
Eq.(\ref{classific equation}) leads to a general
property of the model: {\it the model can be coherent ($\alpha =
constant$) only if there is energy transfer}. Eq.(\ref{classific
equation}) defines the critical energy transfer ${\bf U}_m^{E}$,
which keeps $\alpha$ constant (the coherent solution):
\begin{eqnarray}
{\bf U}^{E}_m =
3\,\frac{\dot a}{a}\,{\cal E}(\beta -w)\,\frac{\alpha}{(\alpha +
1)^2} =
{\bf U}^{E}_m(w, \beta, \gamma, a).
\label{classific equation E}
\end{eqnarray}
${\bf U}^{E}_m$ separates the behaviour of the two-fluid model
with different signs of
$\dot\alpha(a)$: it follows from Eq.(\ref{classific equation})
that
$\dot\alpha(a) > 0$ requires $U_m < {\bf U}_m^{E}$;
$\dot\alpha(a) < 0$ requires $U_m > {\bf U}_m^{E}$.
In other words, the rate $\frac{~{\cal E}_Q}{~{\cal E}_m}$
increases, if $U_m < {\bf U}_m^{E}$ and decreases if $U_m > {\bf
U}_m^{E}$.

In a number of works the energy transfer has been calculated for
a given model of interaction between dust and radiation, i.e. for given
$w(a)$ and $\beta(a)$ (see, for instance,
McIntosh \cite{McIn1967} and \cite{McIn1968}; Sistero \cite{Sis71};
Coley \& Tupper \cite{CT85} and \cite{CT86};
 ).
Now we know that energy transfer
is necessarily required for the coherence.

At the same time Eq.(\ref{classific equation}) gives the condition
for no energy transfer, $U_Q = U_m = 0$, in the form:
\begin{eqnarray}
\alpha = \alpha(1)\,\exp\left(
3\,\int\limits_1^a \frac{\beta - w}{x} dx
\right).
\label{classific equation 11}
\end{eqnarray}
Any other function $\alpha(a)$ leads to non zero energy transfer.

Another form of Eq.(\ref{classific equation}) is obtained by
substituting $\frac{\dot a}{a}$ from Eq.(\ref{0.0 diml}). This
shows how the energy transfer depends on the curvature ($k
\ne 0$):
\begin{eqnarray}
\frac{U_m}{{\cal E}}\,\frac{(\alpha + 1)^2}{\alpha} =
\pm\,3\,\sqrt{{\cal E} - \frac{k}{a^2}}\,
(\beta -w) - \frac{\dot \alpha}{\alpha}.
\label{classific equation 1}
\end{eqnarray}

Using the definition of $\alpha$, Eq.(\ref{alpha}), and its time
derivate,
$\dot\alpha\,(\gamma - w)^2 =
(\gamma - w)\,\dot\beta + (\beta - \gamma)\,\dot w -
(\beta - w)\, \dot\gamma$,
Eq.(\ref{classific equation}) leads to the expression for
the energy transfer:
$$
U_m(\gamma,a) =
\left[
\pm 3\,\frac{(\beta - \gamma)(\gamma - w)}{\beta - w}\right.
$$
\begin{eqnarray}
\left.\mp \frac{a}{(\beta - w)^2}
\biggl(
(\gamma - w)\,\beta^{\prime} + (\beta - \gamma)\,w^{\prime} -
(\beta - w)\,\gamma^{\prime}
\biggr)
\right] \cdot
\label{U_m new}
\end{eqnarray}
$$
{\cal E}(\gamma,a)\,\sqrt{{\cal E}(\gamma,a) - \frac{k}{a^2}},
$$
where ${}^{\prime} = \frac{d}{da}$ and ${\cal E}(\gamma,a)$ is
defined by Eq.(\ref{def en den}). The sign $'+'$ corresponds to
expansion while the sign $'-'$ corresponds to collapse. So, we
found that the energy transfer $U_m$ depends on $\gamma$, $\beta$,
$w$, their  derivates and, also, on the scale factor $a$ (as on an
independent variable) and curvature. For a more realistic situation
with given $k$, $\beta(a)$ and $w(a)$, Eq.(\ref{U_m new})
demonstrates that $\gamma(a)$ is completely defined by $U_m(a)$.

Eqs.(\ref{classific equation}), (\ref{classific equation 1}) and
(\ref{U_m new}) represent in different forms how a convertion of
one-fluid into another (i.e. $U_m \ne 0$) influences the properties
of particular fields (its equations of state, and its energy
transfer rates) and the expansion of space (i.e. $a(t)$).
Eq.(\ref{U_m new}) states the dependence of the energy transfer on
$\gamma$ and scale factor as an independent variable.Excluding
$\gamma$ from Eqs.(\ref{U_m new}), (\ref{def en den}) and
(\ref{motion dl}) we find that energy transfer completely defines
the gravitatng mass as a function of the scale factor and, so, the
dynamics of the universe (see Eq.(\ref{motion dl})).
Therefore, {\it the total gravitating mass
and, consequently, the dynamics of the expansion, is determined by
the energy transfer}.

For the popular flat model Eq.(\ref{classific equation}) becomes:
\begin{eqnarray}
U_m\,\frac{(\alpha + 1)^2}{\alpha} =
3\,~{\cal E}^{3/2}\,(\beta - w) - ~{\cal E}\,\frac{\dot
\alpha}{\alpha}.
\label{classific equation flat}
\end{eqnarray}

The general (ET-NSES) solution for particular fluids is
represented through the function $\alpha(a)$ and two effective
coefficients of the equation of state $\beta(a)$ and $w(a)$:
\begin{eqnarray}
{\cal E}_Q &=& \frac{~{\cal E}\,\alpha}{\alpha + 1},
\qquad\qquad\qquad
~{\cal E}_m = \frac{~{\cal E}}{\alpha + 1},
\nonumber\\
E_Q &=& ~{\cal E}_Q\,a^3,
\qquad\qquad\qquad
E_m = ~{\cal E}_m\,a^3,
\nonumber\\
P_Q &=& w\,~{\cal E}_Q,
\qquad\qquad\qquad
~P_m = \beta\,~{\cal E}_m,
\label{sol 1}
\end{eqnarray}
$$\mu_Q = 3\,\sigma_k(\chi)\,(3\,\beta + 1)\,a^3\,~{\cal E}_Q,$$
$$\mu_m = 3\,\sigma_k(\chi)\,(3\,w + 1)\,a^3\,~{\cal E}_w.$$
Depending on the initial conditions and the class of a model,
there is one (or, possibly, more) moments of time when the
one-fluid model is effectively dust, i.e. $\gamma(a_{ed}) = 0$;
$a_{ed}$ is the ``effective dust scale''. The general expression
for this scale is:
\begin{eqnarray}
\alpha(a_{ed}) = -\frac{\beta}{w}.
\label{a e d g}
\end{eqnarray}

We now show the complicated character of the evolution of
the two-fluid model from initial conditions, defined at
$a_{min}$, to an asymptotical state with $a \to \infty$. To do
this we study the relative contributions of energy, pressure and
gravitating mass of each component to the total energy, total
pressure and total gravitating mass. Eqs.(\ref{sol 1}) give the
criteria for the relative contributions in the form:
\begin{eqnarray}
\alpha &=& \frac{~{\cal E}_Q}{~{\cal E}_m} = \frac{E_Q}{E_m}
\nonumber\\
{\cal P} &=& \frac{|P_Q|}{P_m} =
\frac{\alpha}{\alpha_{\cal P}},
\nonumber\\
{\cal M} &=& \frac{|\mu_Q|}{\mu_m} = \frac{\alpha}{\alpha_{\cal
M}},
\label{def a 1}
\end{eqnarray}
where
\begin{eqnarray}
\alpha_{\cal P} ~&=& \alpha_{\cal P}(a) =
\frac{\beta(a)}{|w(a)|},
\nonumber\\
\alpha_{\cal M} &=& \alpha_{\cal M}(a) = \frac{3\,\beta(a) +
1}{|3\,w(a) + 1|}.
\label{def a 2}
\end{eqnarray}
Three characteristic scales $a_{\cal E}$, $a_{\cal P}$ and $a_{\cal M}$
appear so that
\begin{eqnarray}
a &=& ~a_{\cal E} \quad\mbox{implies}\quad ~~\alpha = 1,
\nonumber\\
a &=& ~a_{\cal P} \quad\mbox{implies}\quad ~{\cal P} = 1, \qquad a_P =
a_{ed},
\nonumber\\
a &=& a_{\cal M} \quad\mbox{implies}\quad {\cal M} = 1.
\label{character scales}
\end{eqnarray}
These are the solutions of the equations:
\begin{eqnarray}
\alpha(a_E) = 1,
\nonumber\\
\frac{|w(a_P)|}{\beta(a_P)}\,\alpha(a_P) = 1,
\nonumber\\ \nonumber\\
\frac{|3\,w(a_{\cal M}) + 1|}{3\,\beta(a_{\cal M}) +
1}\,\alpha(a_{\cal M}) = 1.
\label{character scales 1}
\end{eqnarray}
$\beta(a)$, $w(a)$ and $\alpha(a)$ are model-dependent, so there
is no possibility to study the general case.

Eq.(\ref{classific equation}) shows that the energy transfer is
required
for the coherence ($\alpha = const$). Now we give two other examples of
how to use Eq.(\ref{classific equation}). Substituting  $\alpha =
\alpha_{\cal P}\,{\cal P}$ and $\alpha = \alpha_{\cal M}\,{\cal
M}$ into Eq.(\ref{classific equation}) we find for ${\cal P} =
const$ and ${\cal M} = const$ two critical values  for the energy
transfer:
\begin{eqnarray}
{\bf U}^{P}_m =
\left[
3\,\frac{\dot a}{a}\,(\beta -w) -
\frac{d\,\ln
\alpha_{\cal P}}
{d\tau}
\right]
{\cal E}\frac{{\cal P}\alpha_{\cal P}}{({\cal P}\,\alpha_{\cal P}
+ 1)^2}.
\label{classific equation P}
\end{eqnarray}
and
\begin{eqnarray}
{\bf U}^{M}_m =
\left[
3\,\frac{\dot a}{a}\,(\beta -w) -
\frac{d\,\ln
\alpha_{\cal M}}
{d\tau}
\right]
{\cal E}\frac{{\cal M}\alpha_{\cal M}}{({\cal M}\,\alpha_{\cal M}
+ 1)^2}.
\label{classific equation M}
\end{eqnarray}
$U_m = {\bf U}^{P}_m$ gives $P_Q \sim P_m$ and $U_m = {\bf
U}^{M}_m$ gives $\mu_Q \sim \mu_m$.

\subsubsection{The evolution types}

In this paper we study only the models which are fully dominated
by the matter-component in the very early epoch (i.e. $\alpha < 1$,
${\cal P} < 1$ and ${\cal E} < 1$ ), while fully dominated by
quintessence  at the limit of large time (i.e. $\alpha >
1$, ${\cal P} > 1$ and ${\cal E} > 1$ ). Under this assumption
the characteristic scales $a_E$, $a_P$ and $a_{\mu}$ allow 6
types of nonequalities:
\begin{eqnarray}
&a_{\cal E}& < ~a_{\cal P} < a_{\cal M}, \qquad ~\mbox{EPM}
\nonumber\\
&a_{\cal E}& < a_{\cal M} < ~a_{\cal P}, \qquad ~\mbox{EMP}
\nonumber\\
&a_{\cal P}& < a_{\cal M} < ~a_{\cal E}, \qquad ~\mbox{PME}
\nonumber\\
&a_{\cal P}& < ~a_{\cal E} < a_{\cal M}, \qquad ~\mbox{PEM}
\nonumber\\
&a_{\cal M}& < ~a_{\cal E} < ~a_{\cal P}, \qquad ~\mbox{MEP}
\nonumber\\
&a_{\cal M}& < ~a_{\cal P} < ~a_{\cal E}, \qquad ~\mbox{MPE}.
\label{6 nonequal}
\end{eqnarray}
Each system of nonequalities corresponds to an
``evolution type'', defined by a unique sequence of
the scales $a_E$, $a_P$, and $a_{\mu}$, which represents the
sequence of time periods when the quintessence-component dominates
in total energy, pressure and gravitating mass, respectively. The
ratio between two energies $E_Q$ (actually $E_{\Lambda}$) and
$E_m$ was first considered by Chernin et al. (\cite{CherTerBar00})
to make a division line between the growth and suppression of
structure formation. In Table \ref{epm} the EPM evolution type is
described in detail. Eq.(\ref{classific equation}) leads to a
dependence of $\alpha$ on $U_m$ for given $w(a)$ and $\beta(a)$.
Eqs.(\ref{def a 1}) - (\ref{character scales 1}) lead to a
dependence of the evolution type on $\alpha$ for given $w(a)$ and
$\beta(a)$. So, we resume that for given $w(a)$ and $\beta(a)$ the
evolution type is connected with
the energy transfer. On the other hand, energy transfer defines
the gravitating mass. So, {\it there exists a correspondence
between the dynamics of a model, its evolution type and energy
transfer}.

Below we study how quintessence influences the behavior of the
total gravitating mass, pressure, and the effective equation of
state of the associated one-fluid model and we calculate the
evolution types for a few examples.
\begin{table}
\caption{Detailed description of the EPM evolution type}
\begin{center}
\begin{tabular}{|c|c|} \hline
                              & $~{\cal E}_M > ~{\cal E}_Q$ \\
$a_{min} < a < a_{\cal E}$    & $ P_M > |P_Q| $             \\
                              & $ \mu_M > |\mu_Q| $         \\
                              \hline
                              & $~{\cal E}_M < ~{\cal E}_Q$ \\
$a_{\cal E} < a < a_{\cal P}$ & $ P_M > |P_Q| $        \\
                              & $ \mu_M > |\mu_Q| $         \\
                              \hline
                              & $~{\cal E}_M < ~{\cal E}_Q$ \\
$a_{\cal P} < a < a_{\cal M}$ & $ P_M < |P_Q| $   \\
                              & $ \mu_M > |\mu_Q| $           \\
                              \hline
                              & $~{\cal E}_M < ~{\cal E}_Q$ \\
$a_{\cal M} < a < \infty$     & $ P_M < |P_Q| $          \\
                              & $ \mu_M < |\mu_Q| $           \\
                              \hline
\end{tabular}
\end{center}
\label{epm}
\end{table}

\section{Classifying two-fluid FLRW models: examples}

\subsection{An example of NET-SES models} \label{NET-SES}

This class of two-fluid models has been much studied because the
input equations are simple. In this subsection we study the
evolution types for $w = -1$, $-2/3$, $-1/3$ and for $\beta = 0$,
$1/3$, $2/3$, $1$.

\subsubsection{Input equations and their solutions}

The input equations are given by Eq.(\ref{continuity dimless}) for
every fluid:
\begin{eqnarray}
3\,\frac{\dot a}{a} &=& -\frac{\dot {~{\cal E}_Q}}{ ~{\cal E}_Q +
P_Q}, \quad P_Q = w\,~{\cal E}_Q, \quad w = const.,
\nonumber\\
3\,\frac{\dot a}{a} &=& -\frac{\dot {~{\cal E}_m}}{ ~{\cal E}_m +
P_m}, \quad P_m = \beta\,~{\cal E}_m, \quad \beta = const.,
\nonumber\\
U_Q &=& 0,\qquad U_m = 0. \label{e s 2}
\end{eqnarray}
Eq.(\ref{classific equation}) is written in the form:
\begin{eqnarray}
3\,\frac{\dot a}{a}\,(\beta -w) - \frac{\dot \alpha}{\alpha} = 0
\label{classific equation NETses 1}
\end{eqnarray}
and has a solution
\begin{eqnarray}
\alpha = \alpha(1)\,a^{3\,(\beta - w)}. \label{net-ses alpha}
\end{eqnarray}
The solution of Eqs.(\ref{e s 2}) is well known. Energy density
$~{\cal E}(a)$, characteristic energy content $E(a)$, pressure
$P(a)$ and gravitation mass $\mu(a)$ have the form:
\begin{eqnarray}
{\cal E}_Q(a) &=& \frac{{\cal E}_Q(1)}{a^{3(w + 1)}},\qquad
~~~~~~{\cal E}_m(a) = \frac{{\cal E}_m(1)}{a^{3(\beta + 1)}},
\nonumber\\
E_Q(a) &=& \frac{{\cal E}_Q(1)}{a^{3\,w}},\qquad ~~~~~~~~E_m(a) =
\frac{{\cal E}_m(1)}{a^{3\,\beta}},
\nonumber\\
P_Q(a) &=& w\,\frac{{\cal E}_Q(1)}{a^{3\,(w + 1)}},\qquad
~~~P_m(a) = \beta\,\frac{{\cal E}_m(1)}{a^{3\,(\beta + 1)}},
\label{defs4}
\end{eqnarray}
$$\mu_{Q}(a) = 3\sigma_k(\chi) {\cal E}_Q(1)\,\frac{3\,w +
1}{a^{3\,w}}$$
$$\mu_{m}(a) = 3\sigma_k(\chi) {\cal E}_m(1)\,\frac{3\,\beta +
1}{a^{3\,\beta}}$$. We comment on two models in this
class. First, ($w = - 1/3$, $\beta = 0$) has a constant total
gravitating mass and total scale-dependent energy content. Second,
($w = -1$, $\beta = 0$) produces a total constant pressure and non
constant energy content and gravitating mass (see Table
\ref{twomodels}).

\subsubsection{The condition of coherence}

Assuming coherence (see Eq.(\ref{coh 1}))
Eq.(\ref{net-ses alpha}) leads to
\begin{eqnarray}
\frac{\dot a}{a}\,(\beta - w) = 0, \label{classific equation
netses}
\end{eqnarray}
which for the non-static solution reduces to the simplest case of
one-fluid dust model:
\begin{eqnarray}
\beta = w = 0. \label{classific equation netses dust}
\end{eqnarray}
Assuming asymptotic coherence (Eq.(\ref{coh
2})) Eq.(\ref{net-ses alpha}) gives
\begin{eqnarray}
\alpha^{\prime} = 3\,\alpha\,\frac{\beta - w}{a} =
3\,\alpha(1)\,(\beta - w)\,a^{3\,(\beta -w)\,-1}, \label{as coh
netses}
\end{eqnarray}
so, $\lim_{a \to \infty}\alpha^{\prime} = 0$ only for $\beta - w <
\frac{1}{3}$.

\subsubsection{Asymptotic behaviour and evolution type}

The effective coefficient $\gamma(a)$ of the associated one-fluid
depends on time. Substituting (\ref{net-ses alpha}) into
(\ref{1fm eqs}) one finds $\gamma(a)$:
\begin{equation}
\gamma(a) = \beta - \frac{\beta - w} { 1 + \frac{1}{\alpha(a)} },
\label{gamma net-ses}
\end{equation}
which has the following asymptotical properties:
\begin{eqnarray}
\lim_{a \to 0}{\gamma(a)} = \beta, \qquad \lim_{a \to
\infty}{\gamma(a)} = w, \label{net-ses asimpt 1}
\end{eqnarray}
so, {\it any matter-quintessence NET-SES model is effectively a
matter model at an earlier time and quintessence at the limit of
large time}.

There is a unique moment of time when the total pressure is equal
to zero and the associated one-fluid model is dust-like. The
corresponding effective dust scale $a_{ed}$ is:
\begin{equation}
a_{ed} = \left[ \frac{-w}{\beta}\, \frac{B}{A}
\right]^{\frac{1}{3\,(\beta - w)}}, \label{g n s 1}
\end{equation}

To find the evolution types of NET-SES models, we use the
function $\alpha$ to calculate two functions ${\cal P}$ and
${\cal M}$:
\begin{eqnarray}
\alpha(a) ~~&=& \frac{{\cal E}_Q(1)}{{\cal E}_m(1)}\,a^{3\,(\beta
-
w)}, \nonumber\\
{\cal P}(a) ~&=& \frac{|w|}{\beta}\,\alpha(a),
\nonumber\\
{\cal M}(a) &=& \frac{|3\,w + 1|}{3\,\beta + 1}\,\alpha(a).
\label{defs 1}
\end{eqnarray}
The characteristic scales (see Eq.(\ref{character scales})) may be
calculated analytically:
\begin{eqnarray}
a_{\cal E} ~&=& \left(\frac{{\cal E}_m(1)}{{\cal
E}_Q(1)}\right)^{\frac{1}{3\,(\beta - w)}},
\nonumber\\
a_{\cal P} ~&=& \left(\frac{\beta}{|w|}\frac{{\cal E}_m(1)}{{\cal
E}_Q(1)}\right)^{\frac{1}{3\,(\beta - w)}},
\nonumber\\
a_{\cal M} &=& \left(\frac{3\,\beta+1}{|3\,w+1|}\frac{{\cal
E}_m(1)}{{\cal E}_Q(1)}\right)^{\frac{1}{3\,(\beta - w)}}.
\label{character scales 2}
\end{eqnarray}
\begin{table}
\caption{Two examples of the NET-SES models: $w = -1$ and $\beta =
0$ provide a constant pressure; $w = -1/3$ and $\beta = 0$ provide
a constant pressure. }
\begin{center}
\begin{tabular}{|c|c|c|c|c|c|} \hline
$w$    & $\beta$ & $E$           & $P$           & $\mu$        \\
\hline
$-1$   & $0$     & $A\,a^3  + B$ & $-A$          & $-2\,A\,a^3$ \\
$-1/3$ & $0$     & $A\,a  + B$   & $-A/(3\,a^2)$ & $B$          \\
\hline
\end{tabular}
\end{center}
\label{twomodels}
\end{table}
\begin{table}
\caption{Evolution types for NET-SES models with $w = -1$, $-2/3$,
$-1/3$ and for $\beta = 0$, $1/3$, $2/3$, $1$. These numbers are
obtained from Eqs.(\ref{character scales}) and (\ref{character
scales 1}) with ${\cal E}_Q(1) = 0.7$ and ${\cal E}_m(1) = 0.3$
We denote by $({\bf ME})$ the case of $\alpha_{\cal M} =
\alpha_{\cal E}$ and so on. }
\begin{center}
\begin{tabular}{|c|c|c|c|}   \hline
            & $w=-1$                     & $w=-2/3$                &
$w=-1/3$           \\ \hline $\beta=0$   & $a_E=0.754$
& $a_E=0.655$             &
$a_E=0.429$        \\
            & $a_P=0.00$                 & $a_P=0.00$              &
$a_P=0.00$         \\
            & $a_{\cal M}=0.598$            & $a_{\cal M}=0.655$
& $a_{\cal M}=\infty$   \\
            & ${\bf PME}$                & ${\bf P(ME)}$           &
${\bf PEM}$        \\
            & $a_{ed}=\infty$            & $a_{ed}=\infty$         &
$a_{ed}=\infty$    \\ \hline
$\beta=1/3$ & $a_E=0.809$                & $a_E=0.754$
&
$a_E=0.655$        \\
            & $a_P=0.615$                & $a_P=0.598$             &
$a_P=0.655$        \\
            & $a_{\cal M}=0.809$            & $a_{\cal M}=0.949$
& $a_{\cal M}=\infty$   \\
            & ${\bf P(ME)}$              & ${\bf PEM}$             &
${\bf (PE)M}$      \\
            & $a_{ed}=1.07$              & $a_{ed}=0.950$          &
$a_{ed}=0.655$     \\ \hline
$\beta=2/3$ & $a_E=0.844$                & $a_E=0.809$
&
$a_E=0.754$        \\
            & $a_P=0.778$                & $a_P=0.809$             &
$a_P=0.949$        \\
            & $a_{\cal M}=0.915$            & $a_{\cal M}=1.07$
& $a_{\cal M}=\infty$   \\
            & ${\bf PEM}$                & ${\bf (PE)M}$           &
${\bf EPM}$        \\
            & $a_{ed}=0.915$             & $a_{ed}=0.809$          &
$a_{ed}=0.598$     \\ \hline
$\beta=1$   & $a_E=0.868 $               & $a_E=0.844$
&
$a_E=0.809$        \\
            & $a_P=0.868$                & $a_P=0.915$             &
$a_P=1.07$         \\
            & $a_{\cal M}=0.975$            & $a_{\cal M}=1.11$
& $a_{\cal M}=\infty$   \\
            & ${\bf (PE)M}$              & ${\bf EPM}$             &
${\bf EPM}$        \\
            & $a_{ed}=0.868$             & $a_{ed}=0.778$          &
$a_{ed}=0.615$     \\ \hline
\end{tabular}
\end{center}
\end{table}
The numbers in Table 4 illustrate the complicated
character of the evolution from the matter-dominated model
at the earlier time to the quintessence-dominated stage for large
time. Different physical quantities pass through 'boundaries' ($\alpha = 1$;
 ${\cal P} = 1$; ${\cal M} = 1$) between
the matter-dominated and quintessence-dominated stages at
different times. It is also seen that $a_{ed}$ cannot completely
characterize the evolution type and dynamics of the universe.

\subsubsection{The time dependence of the gravitating mass}

The total gravitating mass in any NET-SES model has its extremum
at
\begin{equation}
a_* = \left( \frac{\beta}{-w}\,\frac{3\,\beta + 1}{3\,w + 1}
\right)^{\frac{1}{3\,(\beta - w)}}. \label{M tot net ses loc}
\end{equation}
As $a_*$ should be real, so $-1 / 3 < w < 0$. The requirement of
positive ${\mu}^{\prime \prime}(a_*) > 0$ is reduced to $\beta >
w$, which is always valid. So, in NET-SES models the gravitating
mass has a minimum for $- 1/3 < w < 0$ and there are no extremums
for $-1 \le w \le -1/3$.

\subsection{A NET-NSES example} \label{NET-NSES}

Here we generalize the known solution for mixed
radiation and dust, first obtained by Chernin (\cite{Chernin65}).
We use it as the matter component of a two-fluid. For quintessence
we assume $w = const.$
A new property, produced by the
complicated  matter-component is that the
effective equation of state is non-stationary. So, this model
is NET-NSES.

\subsubsection{The input equations and their solutions}

The input equations are given by Eqs.(\ref{continuity dimless})
for each fluid:
\begin{eqnarray}
3\,\frac{\dot a}{a} &=& -\frac{\dot {~{\cal E}_Q}}{ ~{\cal E}_Q +
P_Q}, \qquad P_Q = w\,~{\cal E}_Q, \qquad w = const.,
\nonumber\\
3\,\frac{\dot a}{a} &=& -\frac{\dot {{\cal E}_R}}{{\cal E}_R +
P_R}, \qquad ~P_R = \frac{{\cal E}_R}{3},
\nonumber\\
3\,\frac{\dot a}{a} &=& -\frac{\dot {{\cal E}_D}}{{\cal E}_D},
\qquad ~~~~~~~~P_D = 0,
\nonumber\\
{\cal E}_m &=& {\cal E}_R + {\cal E}_D \qquad ~~~~~P_m = P_D + P_R
= P_R,
\nonumber\\
U_Q &=& 0,\qquad U_R = 0,\qquad U_R = 0. \label{e s 2 1}
\end{eqnarray}
The solutions of Eqs.(\ref{e s 2 1}) are well-known. The energy
density $~{\cal E}(a)$, characteristic energy content $E(a)$,
pressure $P(a)$ and gravitating mass $\mu(a)$ have for each
component the form:
\begin{eqnarray}
{\cal E}_Q &=& \frac{{\cal E}_Q(1)}{a^{3(w + 1)}}, \quad
~~~~~~~~{\cal E}_D = \frac{{\cal E}_D(1)}{a^3}, \quad {\cal E}_R =
\frac{{\cal E}_R(1)}{a^4},
\nonumber\\
E_Q &=& \frac{{\cal E}_Q(1)}{a^{3\,w}}, \quad ~~~~~~~~~~E_D =
{\cal E}_D(1), \quad E_R = \frac{{\cal E}_R(1)}{a},
\nonumber\\
P_Q &=& w\,\frac{{\cal E}_Q(1)}{a^{3\,(w + 1)}}, \quad ~~~~~P_D =
0, \quad ~~~~~~P_R = \frac{1}{3}\,\frac{{\cal E}_R(1)}{a^4},
\nonumber\\
\mu_{Q} &=& {\cal E}_Q(1)\,\frac{3\,w + 1}{a^{3\,w}}, \quad
{\mu_D} = {\cal E}_D(1), \quad ~\mu_R = \frac{2\,{\cal
E}_R(1)}{a}. \label{defs33}
\end{eqnarray}
For $\alpha$ and $\beta$ we find:
\begin{equation}
\alpha(a) = \frac{{\cal E}_Q(1)\,a^{1-3\,w}} {{\cal
E}_D(1)\,a+{\cal E}_R(1)}. \label{defs445}
\end{equation}
\begin{equation}
\beta(a) = \frac{~{\cal P}_m}{E_m} = \frac{{\cal E}_R(1)} {{\cal
E}_D(1)\,a+{\cal E}_R(1)}, \label{defs22q}
\end{equation}%
Eq.(\ref{classific equation}) has the form:
\begin{eqnarray}
3\,\frac{\dot a}{a}\,(\beta(a) -w) - \frac{\dot \alpha}{\alpha} =
0. \label{classific equation NETses}
\end{eqnarray}
\subsubsection{The conditions of coherence}

From Eqs.(\ref{coh 2}), (\ref{defs445}) and (\ref{classific equation
NETses}) we find
\begin{eqnarray}
\alpha^{\prime} = \frac{3\,C}{a}\,\frac{a^{1 - 3\,w}}{A\,a + B}.
\label{cond coh netnses 1}
\end{eqnarray}
\begin{eqnarray}
\lim_{a \to \infty}{\alpha^{\prime}} &=&
-3\,w\,\frac{C}{A}\,a^{-(3\,w + 1)} = 0, \quad\mbox{for}\quad w >
-\frac{1}{3}
\nonumber\\
&=& -3\,w\,\frac{C}{A}, \quad~~~~~~~~~~~~~~~~~~\mbox{for}\quad w
=-\frac{1}{3}\nonumber\\
&=& \infty\, ~ ~~~~~~~~~~~~~~~~~~~~~~~~~\quad\mbox{for}\quad w \le
-\frac{1}{3}. \label{cond coh netnses}
\end{eqnarray}

\subsubsection{The asymptotic behaviour and evolution types of
a dust-radiation-quintessence model}

The effective coefficient $\gamma(a)$ of the associated one-fluid
depends on time. Substituting (\ref{defs445}) into
(\ref{1fm eqs}), one finds:
\begin{equation}
\gamma(a) = \frac{ {\cal E}_Q(1)\,w\ + \frac{{\cal
E}_R(1)}{3}\,a^{3\,w - 1} } { {\cal E}_Q(1) + {\cal
E}_D(1)\,a^{3\,w} + {\cal E}_R(1)\,a^{3\,w - 1} }. \label{gamma
net-nses}
\end{equation}
Because $\beta - w > 0$, the associated one-fluid model has the
following asymptotical properties:
\begin{eqnarray}
\lim_{a \to 0}{\gamma(a)} = \frac{1}{3},\qquad \lim_{a \to
\infty}\gamma(a) = w, \label{net-ses asimpt}
\end{eqnarray}
So the one-fluid model, associated with the NET-NSES two-fluid
under consideration is effectively raditive at early
times and quintessence-like at the limit of large time. The model
is effectively dust on the (unique) scale
\begin{equation}
a_{ed} = \left[ \frac{-3\,w\,{\cal E}_Q(1)}{{\cal
E}_R(1)}\right]^{\frac{1}{3\,w - 1}}. \label{g n s}
\end{equation}
The evolution types are characterized by the three functions
$\alpha$, ${\cal P}$ and ${\cal M}$:
\begin{eqnarray}
{\cal P} &=& \frac{3\,w}{{\cal E}_R(1)}\,\left({\cal
E}_D(1)\,a+{\cal E}_R(1)\right)\,\alpha(a),
\nonumber\\
{\cal M} &=& \frac{{\cal E}_Q(1)\,\left(3\,w + 1\right)}{{\cal
E}_D(1)\,a + 2\,{\cal E}_R(1)}\,a^{1 - 3\,w} = \left(3\,w +
1\right)\,\alpha(a). \label{defs12}
\end{eqnarray}
\begin{table}
\caption{These numbers are obtained from Eqs.(\ref{character
scales}) and (\ref{defs12}) with ${\cal E}_Q(1) = 0.7$ and ${\cal
E}_m(1) = 0.3$; we choose here ${\cal E}_D = 0.05$, ${\cal E}_R =
0.25$}
\begin{center}
\begin{tabular}{|c|c|c|}   \hline
$w=-1$                     & $w=-2/3$                & $w=-1/3$
\\ \hline
$a_E=0.803$                & $a_E=0.743$             &
$a_E=0.634$         \\
$a_P=0.587$                & $a_P=0.563$             &
$a_P=0.598$         \\
$a_{\cal M}=0.788$            & $a_{\cal M}=0.921$         &
$a_{\cal M}=\infty$    \\
${\bf PME}$                & ${\bf PEM}$             & ${\bf
PEM}$         \\
$a_{ed}=0.587$             & $a_{ed}=0.563$          &
$a_{ed}=0.598$      \\ \hline
\end{tabular}
\end{center}
\end{table}

In Table 5 three characteristic scales represent two evolution
types for this model: PME and PEM.

We note that the two-fluid description coincides with the
one-fluid model only asymptotically, at $a \to \infty$.

\subsubsection{The time dependence of the gravitating mass}
For $-1/3 < w < 0$ the total gravitating mass has a unique minimum
at a scale $a_*$
\begin{equation}
a_* = \left(\frac{3}{2}\,\frac{{\cal E}_Q(1)}{{\cal E}_R(1)}\,
(-w)\,(3\,w+ 1) \right)^{\frac{1}{3w -1}}. \label{M tot net ses
loc e}
\end{equation}

\subsection{An ET-SES example} \label{ET-SES}

Unlike the NET-SES models, ET-SES models allow
coherence. We inspect here an ET-SES example, with the
two-fluids described by constant $\beta$ and
$w$.

An important class of dark energy (DE) models is the so-called
coupled quintessence (see e.g. Wetterich \cite{Wett95}; Amendola
\cite{amendola99}), where the total (matter + DE) energy momentum
tensor is conserved, whereas in ordinary quintessence
the matter and DE are separately conserved. For coupled
quintessence, a stationary DE model was recently found (Amendola
\& Tocchini-Valentini \cite{amendola00}) which predicts a
coherent behaviour of matter and DE densities at late cosmic
epochs. Such a relation was studied by Baryshev et al.
(\cite{baryshev0011528}, \cite{BarCherTer01}) in order to solve
the problem of the low velocity dispersion in the local Hubble
flow.

The coherence ($\alpha = constant$) implies $\gamma = constant$
and the model is greatly simplified. All these properties are
produced by the energy transfer.

\subsubsection{Input equations and their solution} \label{iets}

The associated one-fluid model is described by
\begin{eqnarray}
3\,\frac{\dot a}{a} = -\frac{\dot {~{\cal E}}}{ ~{\cal E} + P},
\quad P = \gamma\,~{\cal E}, \quad \gamma = const. \label{4.5.1
1}
\end{eqnarray}
which have the solution:
\begin{eqnarray}
{\cal E} &=& \frac{~{\cal E}(1)}{\,a^{3\,(\gamma + 1)}},\qquad E =
\frac{~{\cal E}(1)}{a^{3\,\gamma}},
\nonumber\\
P &=& \frac{~{\cal E}(1)\,\gamma}{\,a^{3\,(\gamma + 1)}}\qquad
~~\mu = \frac{~{\cal E}(1)\,(3\,\gamma + 1)}{a^{3\,\gamma}}.
\label{et-ses 1}
\end{eqnarray}
For the quintessence component we find:
\begin{eqnarray}
~{\cal E}_Q &=& \frac{~{\cal E}(1)}{a^{3\,(\gamma +1)}}\,
\frac{\alpha}{\alpha + 1}, \quad E_Q = \frac{~{\cal
E}(1)}{a^{3\,\gamma}}\,
\frac{\alpha}{\alpha + 1}, \nonumber\\
P_Q &=& \frac{~{\cal E}(1)\,w}{a^{3\,(\gamma +1)}}\,
\frac{\alpha}{\alpha + 1}, \label{et-ses 2 1}
\end{eqnarray}
$$\mu_Q = 3\sigma_k(\chi) ~{\cal E}(1)\,\frac{\alpha}{\alpha + 1}\,
\frac{3\,w + 1}{a^{3\,\gamma}}$$
and for the matter-component:
\begin{eqnarray}
~{\cal E}_m &=& \frac{~{\cal E}(1)}{a^{3\,(\gamma +1)}}\,
\frac{1}{\alpha + 1}, \quad E_m = \frac{~{\cal
E}(1)}{a^{3\,\gamma}}\,
\frac{1}{\alpha + 1}, \nonumber\\
P_m &=& \frac{~{\cal E}(1)\,\beta}{a^{3\,(\gamma +1)}}\,
\frac{1}{\alpha + 1}, \label{et-ses 2}
\end{eqnarray}
$$\mu_m = 3\sigma_k(\chi) ~{\cal E}(1)\,\frac{1}{\alpha + 1}\,
\frac{3\,\beta + 1}{a^{3\,\gamma}}$$
From Eq.(\ref{classific equation}) we find the energy transfer
\begin{eqnarray}
U_m = 3\,\frac{\alpha}{(\alpha + 1)^2}\,\frac{{\cal
E}(1)}{a^{3(\gamma + 1)}}\,\frac{\dot a}{a}\,(\beta - w).
\label{U_m et-ses}
\end{eqnarray}

\subsubsection{The condition of coherence}

The coherence implicitly exists in this model.

\subsubsection{The asymptotic behaviour and evolution types}

The asymptotic behaviour strongly depends on the sign and value of
$\gamma$, obtained at the end of the section
(\ref{iets}).

Three functions, generally defining the evolution type,
\begin{eqnarray}
\alpha &=& \frac{{\cal E}_Q}{{\cal E}_m} = constant,\nonumber\\
{\cal P} &=& \frac{|w|}{\beta}\,\alpha = constant,\nonumber\\
{\cal M} &=& \frac{|3\,w + 1|}{3\,\beta + 1}\,\alpha = constant.
\label{e t et-ses}
\end{eqnarray}
are constant in this model (a fixed ratio between the
energy, pressure and gravitating masses of the quintessence and
matter components). What component dominates during all the
evolution, depends on the value of $\alpha$.

Let us study now the case of the Amendola \& Tocchini-Valentini
(\cite{amendola00}) solution, which is reached at late times: $w =
-0.7$, $\beta = 0$, $\alpha = 7/3$. The equation of state of the
associated one-fluid model is defined by
\begin{eqnarray}
\gamma = \frac{w\,\alpha + \beta}{\alpha + 1} = -\frac{49}{100}
\approx - \frac{1}{2}, \label{am gamma}
\end{eqnarray}
which gives
\begin{eqnarray}
{\cal E} &=& \frac{{\cal E}(1)}{a^{153/100}}
\approx \frac{{\cal E}(1)}{a^{3/2}}\nonumber\\ \nonumber\\
E &=& {\cal E}(1)\,a^{147/100}
\approx {\cal E}(1)\,a^{3/2}\nonumber\\ \nonumber\\
{\cal P} &=& -\frac{49}{100}\,\frac{{\cal E}(1)}{a^{153/100}}
\approx -\frac{1}{2}\,\frac{{\cal E}(1)}{a^{3/2}} \nonumber\\
\nonumber\\
\mu &=& - \frac{47}{100}\,{\cal E}(1)\,a^{147/100} \approx -
\frac{1}{2}\,{\cal E}(1)\,a^{3/2}. \label{am func}
\end{eqnarray}
The energy transfer for the flat ($k = 0$) model is:
\begin{eqnarray}
U_m = 3\,\frac{\alpha}{(\alpha + 1)^2}\, \frac{{\cal
E}^{3/2}(1)}{a^{9/2\,(\gamma + 1)}}\,(\beta - w) \label{U_m et-ses
A}
\end{eqnarray}
This gives for the Amendola \& Tocchini-Valentini
 solution
\begin{eqnarray}
U_m = \frac{441}{1000}\,\frac{{\cal E}^{3/2}(1)}{a^{459/200}}
\approx \frac{2}{5}\,\frac{{\cal E}^{3/2}(1)}{a^{2.3}}. \label{U_m
et-ses A 1}
\end{eqnarray}
Thus this model is asymptotically NET-SES.

\section{Conclusions}

The cosmological view that the
universe is described by a FLRW model with $\Omega_m^0 \approx
0.3$, $\Omega_{\Lambda}^0 \approx 0.7$, and $w \leq -1/3$ has
initiated many studies
of FLRW models with  an essential
$\Lambda$ component at late epochs.
Usually one has viewed $\Lambda$ and matter as
independent substances so that
the energy-momentum tensors of the partial fluids are
separately conserved (Eq.(\ref{divergence 12})).
However, there are a number of suggestions in the litterature
on the particular forms of energy transfer between dark energy
and matter.
This has motivated us to consider on a phenomenological level the general case
when one-fluid converts into another and the equation of state
for both components is non-stationary (Section 3).
 The properties of the model
we  gave in terms of the coefficients of the equation of
state for two partial models ($\beta$ for a component with
positive pressure and $w$ for a negative pressure one)
and a coefficient $\gamma$ for the associated
one-fluid. The energy transfer, when a
one-fluid converts into another, was represented via these
coefficients and the scale factor (Eqs.(\ref{classific
equation}) - (\ref{classific equation flat})).

 We have analyzed four classes of models defined by
the presence or absence of the energy transfer
and by the stationarity ($w = const.$, and $\beta = const.$)
or/and non-stationarity ($w$ or $\beta$ time
dependent) of the equations of state (see Tables 1, 4, 5).
It was shown in sect.3 that

\begin{itemize}

\item[*]{for given $w$ and $\beta$, the energy
transfer defines $\gamma$ and, therefore, the total gravitating
mass and the dynamics of the model.}
\item[*]{also, the model can be coherent only if there is energy transfer.}

\end{itemize}
The classification was illustrated with interesting examples of
two-fluid FLRW models in
sections 4.1 -- 4.3.

From the behaviour of the energy content,
gravitating mass, and pressure as functions of the scale factor
we have defined
three characteristic scales, $a_E$, $a_{\cal P}$ and $a_{\cal M}$.
These separate time intervals when quintessence energy, pressure and
gravitating mass were dominating (Eqs.(\ref{def a 1}) -
(\ref{character scales 1})). Any sequence of the scales defines
one of 6 evolution types of the model (Eqs.(\ref{6 nonequal})).
There is a correspondence between the dynamics of a model, its evolution type
and energy transfer.

\section{Acknowledgements}

We thank A. Grib, V. Dorofeev and S. Krasnikov for useful
discussions and A. Coley for the list of references. This work has been
supported by the Academy of Finland (the project: ``Cosmology from the
local to the deep galaxy universe'').

{}

\end{document}